\documentclass[10pt, twocolumn, pre, aps, amsmath, superscriptaddress]{revtex4}
\usepackage{amsmath, graphicx, subfigure, float, hhline, multirow}
\begin{document}
    \title{Successively Thresholded Domain Boundary Roughening Driven by Pinning Centers and Missing Bonds: Hard-Spin Mean-Field Theory Applied to d=3 Ising Magnets}
    \author{Tolga \c{C}a\u{g}lar}
    \affiliation{Faculty of Engineering and Natural Sciences, Sabanc\i\ University, Tuzla, Istanbul 34956, Turkey}
    \author{A. Nihat Berker}
    \affiliation{Faculty of Engineering and Natural Sciences, Sabanc\i\ University, Tuzla, Istanbul 34956, Turkey}
    \affiliation{Department of Physics, Massachusetts Institute of Technology, Cambridge, Massachusetts 02139, USA}
    \begin{abstract}
Hard-spin mean-field theory has recently been applied to Ising
magnets, correctly yielding the absence and presence of an interface
roughening transition respectively in $d=2$ and $d=3$ dimensions and
producing the ordering-roughening phase diagram for isotropic and
anisotropic systems. The approach has now been extended to the
effects of quenched random pinning centers and missing bonds on the
interface of isotropic and anisotropic Ising models in $d=3$. We
find that these frozen impurities cause domain boundary roughening
that exhibits consecutive thresholding transitions as a function
interaction of anisotropy. For both missing-bond and pinning-center
impurities, for moderately large values the anisotropy, the systems
saturate to the "solid-on-solid" limit, exhibiting a single
universal curve for the domain boundary width as a function of
impurity concentration.


%05.50.+q    Lattice theory and statistics (Ising, Potts, etc.) (see
%also 64.60.Cn Order-disorder transformations, and 75.10.Hk Classical
%spin models)

%05.70.Np    Interface and surface thermodynamics (see also 68.35.Md
%Surface thermodynamics, surface energies in surfaces and interfaces)

%64.60.De    Statistical mechanics of model systems (Ising model,
%Potts model, field-theory models, Monte Carlo techniques, etc.)

%68.35.Dv    Composition, segregation; defects and impurities

%75.60.Ch    Domain walls and domain structure (for magnetic bubbles
%and vortices, see 75.70.Kw)

\end{abstract}
\maketitle

\section{Introduction}

Hard-spin mean-field theory \cite{HSMFT01,HSMFT02} has recently been
applied to Ising magnets, correctly yielding the absence and
presence of an interface roughening transition respectively in $d=2$
and $d=3$ dimensions and producing the ordering-roughening phase
diagram for isotropic and anisotropic systems.\cite{HSMFT16} The
approach is now extended to the effects of quenched random pinning
centers and missing bonds on the interface of isotropic and
uniaxially anisotropic Ising models in $d=3$. We find that these
frozen impurities cause domain boundary roughening that exhibits
consecutive thresholding transitions as a function interaction of
anisotropy.  We also find that, for both missing-bond and
pinning-center impurities, for moderately large values the
anisotropy, the systems saturate to the "solid-on-solid" limit,
exhibiting a single universal curve for the domain boundary width as
a function of impurity concentration.

\begin{figure}[th!]
\centering
\includegraphics[scale=1]{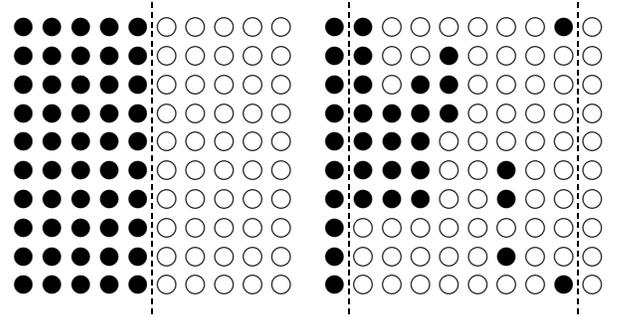}
\caption{A $yz$ plane at temperature $1/J_{xy}=0.1$. Filled and
empty circles respectively represent the calculated local
magnetizations with $m_i>0$ and $m_i<0$. The left side is for the
pure system, $p=0$. The right side is calculated with quenched
random pinning centers with concentration $p=0.24$. Islands that are
disconnected from the pinned $z$ boundary plane of their own sign
(typically occurring around an opposite pinning center deep inside a
bulk phase) do not enter the interface width calculation and are not
shown here. Thus, the disconnected pieces seen in this figure are
actually part of an overhang, connected to the corresponding $z$
boundary plane via the other $yz$ planes. The dashed lines delimit
the domain boundary and the separation between these dashed lines
gives the domain boundary width in this $yz$ plane. The same
procedure for determining the interface width is also applied to the
missing bond systems.}
\end{figure}

\begin{figure}[!ht]
\centering
\includegraphics[scale=1]{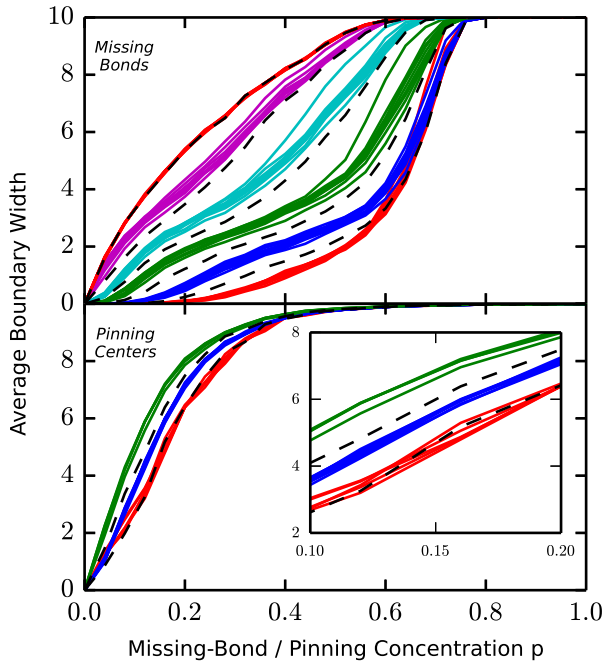}
\caption{(Color online) Calculated domain boundary widths versus
impurity concentration $p$ for different anisotropy $J_z/J_{xy}$
values, at temperature $1/J_{xy}=0.1$.  In the upper panel, the
horizontal axis $p$ is the ratio of the number of missing bonds to
the to the total number of bonds when none is missing.  In the lower
panel, the horizontal axis $p$ is the ratio of the number of pinned
sites to the total number of sites.  In the upper panel for missing
bonds, from the bottom to the top curves, the anisotropies are
$J_z/J_{xy} = 0.1$ to $5.0$ with $0.1$ intervals and $J_z/J_{xy} =
5.5$ to $10$ with $0.5$ intervals. The dashed curves are calculated
with the predicted threshold anisotropy values of $J_z/J_{xy} =
1,2,3,4,5$. In the lower panel for pinning centers, the anisotropies
are $J_z/J_{xy} = 0.5$ to $2.5$ with $0.1$ intervals. The dashed
curves are calculated with the predicted threshold anisotropy values
of $J_z/J_{xy} = 1,2$. Beyond $J_z/J_{xy} \simeq 5$ and $2.3$,
respectively for missing bonds and pinning centers, the system
saturates to the $J_z/J_{xy} \rightarrow \infty$ "solid-on-solid"
limit, exhibiting a single universal curve for the domain boundary
width as a function of impurity concentration, for all $J_z/J_{xy}
\gtrsim 5$ and $J_z/J_{xy} \gtrsim 2.3$ respectively.}
\end{figure}

\begin{figure}[!ht]
\centering
\includegraphics[scale=0.95]{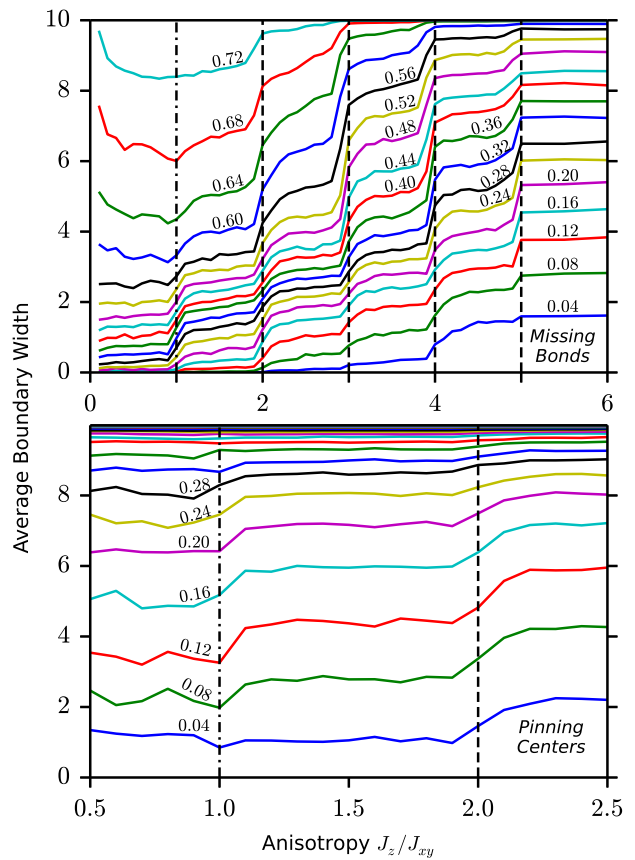}
\caption{(Color online) Calculated domain boundary widths versus
anisotropy $J_z/J_{xy}$, at temperature $1/J_{xy}=0.1$.  The
consecutive curves, bottom to top, are for impurity concentration
values of $p = 0.04$ to $0.72$ (top panel) and $1$ (bottom panel)
with $0.04$ intervals.  These values of $p$ are noted next to the
curves. In the upper panel, $p$ is the ratio of the number of
missing bonds to the to the total number of bonds when none is
missing.  In the lower panel, $p$ is the ratio of the number of
pinned sites to the total number of sites. The curves show the
deviations from the isotropic case $J_z/J_{xy}=1$ (vertical
dash-dotted line) in the directions of strongly coupled planes
$J_z/J_{xy}>1$ or weakly coupled planes $J_z/J_{xy}<1$. The
predicted threshold values are shown with the vertical dash-dotted
and dashed lines and are well reproduced by the calculated widths.
It is clearly seen to the right of this figure that beyond
$J_z/J_{xy} \simeq 5$ and $2.3$, respectively for missing bonds and
pinning centers, the system saturates to the $J_z/J_{xy} \rightarrow
\infty$ "solid-on-solid" limit, exhibiting a single universal value
for the domain boundary width as a function of impurity
concentration, for all $J_z/J_{xy} \gtrsim 5$ and $J_z/J_{xy}
\gtrsim 2.3$ respectively.} \label{fig:aw}
\end{figure}

\section{The Anisotropic $d=3$ Ising Model with Impurities and Hard-Spin Mean-Field Theory}

\subsection{The \texttt{d=3} Anisotropic Ising Model}

The $d=3$ anisotropic Ising model is defined by the Hamiltonian
        \begin{equation}
            \centering
            \label{eq:hamil}
            - \beta {\cal H} = J_{xy} \sum_{\left<ij\right>}^{xy}{s_is_j} + J_z \sum_{\left<ij\right>}^{z}{s_is_j},
        \end{equation}
where at each site $i$ of a cubic lattice, the spin takes on the
values $s_i = \pm 1$. The first sum is over the nearest-neighbor
pairs of sites along the $x$ and $y$ spatial directions and the
second sum is over the nearest-neighbor pairs of sites along the $z$
spatial direction. The system has ferromagnetic interactions
$J_{xy}, J_z > 0$, periodic boundary conditions in the $x$ and $y$
directions and oppositely fixed boundary conditions at the two
terminal planes in the $z$ spatial direction, which yields a domain
boundary within the system when in the ordered phase.  Thus, the
system is generally uniaxially anisotropic.  We systematically study
the anisotropic $J_{xy} \neq J_z$ as well as the isotropic $J_{xy} =
J_z$ cases.

\subsection{Method: Hard-Spin Mean-Field Theory}

In our current study, hard-spin mean-field theory
\cite{HSMFT01,HSMFT02}, which has been qualitatively and
quantitatively successful in frustrated and unfrustrated,
equilibrium and non-equilibrium magnetic ordering problems
\cite{HSMFT03,HSMFT04,HSMFT05,HSMFT06,HSMFT07,HSMFT08,HSMFT09,HSMFT10,HSMFT11,HSMFT12,HSMFT13,HSMFT14,HSMFT15,HSMFT16,HSMFT17},
including recently the interface roughening transition
\cite{HSMFT16}, is used to study the roughening of an interface by
quenched random pinning center sites or missing bonds. The
self-consistency equations of hard-spin mean-field theory
\cite{HSMFT02} are
        \begin{equation}
            \centering
            \label{eq:update}
            m_i = \sum_{\left\{s_j\right\}}\left[\left(\prod_j{\frac{1 + m_js_j}{2}}\right)
            \tanh\left({\sum_jJ_{ij}s_j}\right)\right],
        \end{equation}
where $m_i=<s_i>$ is the local magnetization at site $i$, the sum
$\{s_j\}$ is over all possible values of the spins $s_j$ at the
nearest-neighbor sites $j$ to site $i$, and $m_j$ are the
magnetizations at the nearest-neighbor sites.  These coupled
equations for all sites are solved by local numerical iteration, in
a $10\times10\times10$ system.

\section{Domain Boundary Widths}

\subsection{Determination of the Domain Boundary Width}

In our study, the domain boundary is roughened in two ways: (1)
Magnetic impurities are included in the system by pinning randomly
chosen sites to $s_i = + 1$ or to $s_i = - 1$. The impurity
concentration $p$ in this case is the ratio of the number of pinned
sites to the total number of sites. The numbers of $+1$ and $-1$
pinned sites are fixed to be equal, to give both domains an equal
chance to advance over its counter. (2) Missing bonds are created by
removing randomly chosen bonds. In this case, the concentration $p$
is given by the ratio of the number of removed bonds to the total
number of bonds when none is missing.

The domain boundary width is calculated by first considering each
$yz$ plane.  The boundary width in each $yz$ plane is calculated by
counting the number of sites, in the $z$ direction, between the two
furthest opposite magnetizations in the plane (Fig. 1).  This number
is averaged over all the $yz$ planes.  The result is then averaged
over $100$ independent realizations of the quenched randomness. We
have checked that our results are robust with respect to varying the
number of independent realizations of the quenched randomness, as
shown below.

\subsection{Impurity Effects on\\
the Domain Boundary Width}

Our calculated domain boundary widths, as a function of impurity
(\textit{i.e.}, missing-bond or pinned-site) concentration $p$ at
temperature $1/J_{xy}=0.1$, are shown in Fig. 2. The different
curves are for different interaction anisotropies $J_z/J_{xy}$. In
the lower panel for pinning-center impurity, the domain boundary
roughens with the introduction of infinitesimal impurity, for all
anisotropies: The curves have finite slope at the pure system. In
the upper panel for missing-bond impurity, the domain boundary
roughens with the introduction of infinitesimal impurity for
strongly coupled planes $J_z/J_{xy} > 2.5$, whereas for weakly
coupled planes $J_z/J_{xy} < 2.5$, it is seen that infinitesimal or
small impurity has essentially no effect on the flat domain
boundary.  In the latter cases, the curves reach the pure system
with zero slope.

For both missing-bond and pinning-center impurities, for moderately
large values of $J_z/J_{xy}$, we find (Figs. 2 and 3) that the
systems saturate to the $J_z/J_{xy} \rightarrow \infty$
"solid-on-solid" limit \cite{sons}. Thus, the systems exhibit a
single universal curve for the domain boundary width as a function
of impurity concentration, onwards from all moderately large values
of $J_z/J_{xy}$.

\begin{figure}[!ht]
\centering
\includegraphics[scale=1]{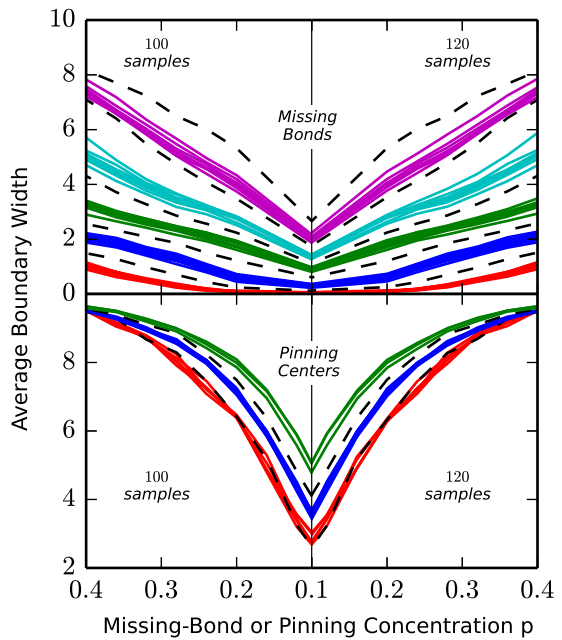}
\caption{(Color online) Calculated domain boundary widths versus
impurity concentration $p$ for different anisotropy $J_z/J_{xy}$
values, at temperature $1/J_{xy}=0.1$.  These curves are obtained by
averaging over $100$ (left panels) and $120$ (right panels)
independent realizations of the quenched randomness.  In the upper
panel, the horizontal axis $p$ is the ratio of the number of missing
bonds to the to the total number of bonds when none is missing.  In
the lower panel, the horizontal axis $p$ is the ratio of the number
of pinned sites to the total number of sites.  In the upper panel
for missing bonds, from the bottom to the top curves, the
anisotropies are $J_z/J_{xy} = 0.1$ to $5.0$ with $0.1$ intervals.
The dashed curves are calculated with the predicted threshold
anisotropy values of $J_z/J_{xy} = 1,2,3,4,5$. In the lower panel
for pinning centers, the anisotropies are $J_z/J_{xy} = 0.5$ to
$2.3$ with $0.1$ intervals. The dashed curves are calculated with
the predicted threshold anisotropy values of $J_z/J_{xy} = 1,2$.
Comparison of the left and right panels shows that our results are
robust with respect to varying the number of independent
realizations of the quenched randomness.}
\end{figure}

\begin{figure}[!ht]
\centering
\includegraphics[scale=1]{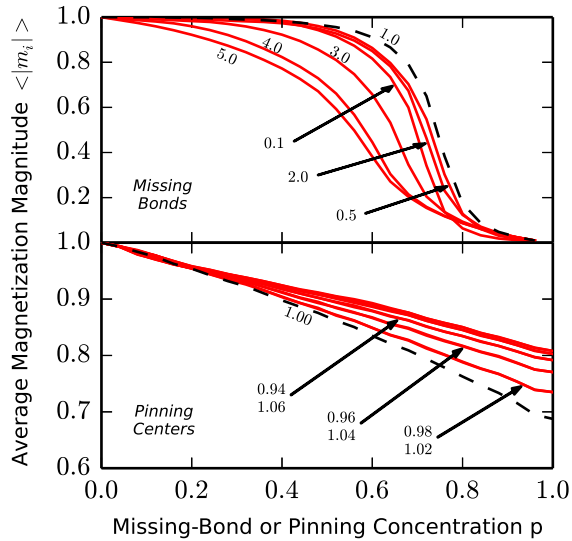}
\caption{(Color online) Calculated local magnetization magnitudes
$<|m_i|>$ averaged across the system versus impurity concentration
$p$ for different anisotropy $J_z/J_{xy}$ values, at temperature
$1/J_{xy}=0.1$.  In the upper panel, the horizontal axis $p$ is the
ratio of the number of missing bonds to the to the total number of
bonds when none is missing.  In the lower panel, the horizontal axis
$p$ is the ratio of the number of pinned sites to the total number
of sites. In each panel, the dashed curve corresponds to the
isotropic case $J_z/J_{xy}=1$.  The full curves are for the
anisotropic cases.  Some of the $J_z/J_{xy}$ values for the
anisotropic cases are indicated next to the corresponding curves.
Note that the average magnetization magnitude curve of the isotropic
case constitutes an upper boundary to the curves of the anisotropic
cases for the missing bonds system (upper panel). The average
magnetization magnitude curve of the isotropic case constitutes a
lower boundary to the curves of the anisotropic cases for the
pinning center system (lower panel). This is understandable by the
fact that missing bonds weaken the connectivity and therefore the
magnetization of the system, whereas pinning centers constitute a
strong aligning field to their neighboring spins.  In curves in the
lower panel, the deviation from the isotropic case is symmetric, so
that each curve corresponds to two values of the anisotropy
$J_z/J_{xy}$ which are above and below the isotropic case
$J_z/J_{xy}=1$.} \label{fig:aw}
\end{figure}

\subsection{Successive Roughening Thresholds}

A bunching of the curves is visible, in the domain-boundary width
curves in Fig. 2, especially in the upper panel for missing-bond
impurity.  This corresponds to a thresholded domain boundary
roughening, controlled by the interaction anisotropy. This effect is
also visible in Fig. 3, by the steps in the curves which give the
domain boundary widths as a function of the interaction anisotropy
$J_z/J_{xy}$ for different impurity concentrations $p$, at
temperature $1/J_{xy}=0.1$. We have checked that our results are
robust with respect to varying the number of independent
realizations of the quenched randomness. This is shown in Fig. 4.

Thresholded domain boundary roughening can be understood by
considering the effect of increasing the anisotropy.  We first
discuss the case of missing-bond impurity. Upon increasing $J_z$,
for what value of $J_z$ will a spin flip, \textit{e.g.}, from +1 to
-1, thereby increasing the domain boundary width (directly and/or by
inducing a flip cascade)? Increasing $J_z$ can flip a spin and
increase the width only if one of its bonds in the $\pm z$ direction
is missing and the non-missing bond connects to a -1 spin. This flip
will then happen for $J_z = (q-q') J_{xy}$, where $(q,q')$ are the
numbers of $xy$ neighbors bonded to the flipping spin that are
respectively +1, -1. The possible values are $(q,q') =
(4,0),(3,0),(2,0),(1,0),(3,1),(2,1)$, giving the threshold values of
$J_z/J_{xy} = 1,2,3,4$, in fact calculationally seen in the top
panels of Figs. 2 and 3.  Furthermore, the simultaneous flip of two
neighboring spins gives the threshold value of $J_z/J_{xy} = 5$,
also calculationally seen in the top panels of Figs. 2 and 3. Beyond
$J_z/J_{xy} = 5$, the system saturates to the $J_z/J_{xy}
\rightarrow \infty$ "solid-on-solid" limit \cite{sons}, exhibiting a
single universal curve for the domain boundary width as a function
of impurity concentration, for all $J_z/J_{xy} \gtrsim 5$.

We now discuss the case of pinned-site impurity. We again consider
the effect of increasing $J_z$ and investigate the value of $J_z$
that will flip the spin, \textit{e.g.}, from +1 to -1, thereby
increasing the domain boundary width (again, directly and/or by
inducing a flip cascade).  Increasing $J_z$ can flip this spin only
if both of its neighbors in the $\pm z$ direction are -1, with one
of these being part of a disconnected island seeded by a pinning
center. This flip will then happen for $2 J_z = (q-q') J_{xy}$,
where again $q$ and $q'$ are the numbers of $xy$ neighbors bonded to
the flipping spin that are respectively +1 and -1. The possible
values are $(q,q') = (4,0),(3,1)$, giving the threshold values of
$J_z/J_{xy} = 1,2$, calculationally seen in the bottom panels of
Figs. 2 and 3.  Beyond $J_z/J_{xy} \simeq 2.3$, the system saturates
to the $J_z/J_{xy} \rightarrow \infty$ "solid-on-solid" limit
\cite{sons}, exhibiting a single universal curve for the domain
boundary width as a function of impurity concentration, for all
$J_z/J_{xy} \gtrsim 2.3$.

On a similar vein, in the limit of $xy$ planes weakly coupled due to
low $J_z/J_{xy}$ and high concentration of missing bonds, the domain
boundary gains by the intermediacy of sending overhangs in the
lateral $x$ and $y$ directions, eventually covering the whole system
via randomly magnetized $xy$ planes. In this case, the spin is
flipped by the effect of $J_{xy}$ upon decreasing $J_z$. This flip
occurs at $2 J_z = (q-q') J_{xy}$, where $(q,q')$ has to be such
that $J_z/J_{xy}$ is low.  Thus, $(q,q') = (2,1)$. (Other pairs of
values, (3,0) and (1,0) do not contribute to this spread of
overhangs.)  Indeed, in Fig. 3, a rise in the domain for decreasing
$J_z<0.5$ is seen at high missing bond concentration.

The curves in Fig. 3 are domain boundary widths that are affected by
complicated (due to the random geometric distribution of the
impurities) cascades of flips of groups of spins, occurring
continuously as the interaction anisotropy is changed. The arguments
given above are for single-spin flips, which strongly affect the
boundary width at the specific anisotropy ratios.

We note that since in this system the interactions acting on a given
spin $s_i$ can be competing, due to the presence of the interface or
of a neighboring pinning center, all of the local magnetizations
$m_i = <s_i>$, where the averaging is thermal, are not saturated
even at low temperatures. Such an effect has been seen down to zero
temperature in other systems with competing interactions, as for
example shown in Fig. 3 of Ref. \cite{Yesilleten}.  In our present
study, the calculated magnitudes of the local magnetizations
averaged across our current system, $<|m_i|>$, are given in Fig. 5
and show this unsaturation.

\section{Conclusion}

The effects of quenched random pinning centers and missing bonds on
the interface of isotropic and uniaxially anisotropic Ising models
in $d=3$ have been investigated by hard-spin mean-field theory. We
find that the frozen impurities cause domain boundary roughening
that exhibits consecutive thresholding transitions as a function
interaction of anisotropy $J_z/J_{xy}$.  The numerical results,
showing the thresholding transitions as the bunching of domain
boundary width versus impurity concentration curves (Fig. 2) and
steps in the domain boundary width versus anisotropy curves (Fig. 3)
agree with our spin-flip arguments at the interface. The threshold
effect should be fully observable in experimental magnetic samples
with good crystal structure and point impurities. For both
missing-bond and pinning-center impurities, for moderately large
values of $J_z/J_{xy}$, the systems saturate to the $J_z/J_{xy}
\rightarrow \infty$ "solid-on-solid" limit, thus exhibiting a single
universal curve for the domain boundary width as a function of
impurity concentration, onwards from all moderately large values of
$J_z/J_{xy}$.

\begin{acknowledgments}
Support by the Alexander von Humboldt Foundation, the Scientific and
Technological Research Council of Turkey (T\"UBITAK), and the
Academy of Sciences of Turkey (T\"UBA) is gratefully acknowledged.
\end{acknowledgments}

\end{document}